\documentstyle[lprocl]{article}

\input{psfig}

\def\Journal#1#2#3#4{{#1} {\bf #2}, #3 (#4)}

\def\NIMA{{\em Nucl. Instrum. Methods} A}
\def\NPB{{\em Nucl. Phys.} B}
\def\PLB{{\em Phys. Lett.}  B}
\def\PRL{\em Phys. Rev. Lett.}
\def\PRD{{\em Phys. Rev.} D}
\def\ZPC{{\em Z. Phys.} C}

\def\JETP{\em Sov. Phys. JETP}
\def\SNP{\em Sov. J. Nucl. Phys.}

\def\PREV{\em Phys. Rev.}

\def\be{\begin{equation}}
\def\ee{\end{equation}}
\def\bea{\begin{eqnarray}}
\def\eea{\end{eqnarray}}

\begin{document}

\begin{titlepage}
 
\vspace*{5mm} 
\begin{large}
\begin{flushleft}
{\tt DESY 97-248    \hfill    ISSN 0418-9833} \\
{\tt December 1997}    
\end{flushleft}
\end{large} 
\vspace*{20mm}         
\begin{center}
\setlength{\baselineskip}{1.333\baselineskip} 
{\Large \bf THE STRUCTURE OF HADRONS}
                       
\vspace*{1.0cm}
{\large  Vladimir CHEKELIAN (SHEKELYAN)} \\
\vspace*{0.3cm}
ITEP (Moscow)\\E-mail: shekeln@mail.desy.de
\vspace*{2cm}\\
\end{center}
\begin{center}
\bf{Abstract}
\end{center}
Recent experimental results contributing to the understanding of the 
structure of the nucleon are reviewed.
They include the final NMC ($\mu$$N\rightarrow\mu$$X$) results
on proton and deuteron structure functions ;
a re-analysis of the CCFR ($\nu$Fe$\rightarrow lX$) data on $F_2$
and $xF_3$; new preliminary results from
CDF on charge asymmetry in $W$ production,
from E866 on Drell-Yan $\mu$-pair production and from
E706 on prompt photon production.
New results from HERA on $F_2$, on the gluon density at low $x$, 
on the charm contribution 
$F_2^{c\overline c}$, 
on a determination of $F_L$ 
and on measurements in the very low $Q^2$ region
are discussed.

\vspace*{6cm}
\begin{center}
\it{ Invited talk at the 18th International Symposium on Lepton-Photon 
Interactions, \\
Hamburg, July 1997}
\end{center} 

\end{titlepage}
%
%
%
\title{THE STRUCTURE OF HADRONS}

\author{Vladimir CHEKELIAN (SHEKELYAN)}

\address{ITEP (Moscow)\\E-mail: shekeln@mail.desy.de} 


\maketitle\abstracts{ 
Recent experimental results contributing to the understanding of the 
structure of the nucleon are reviewed.
They include the final NMC ($\mu$$N\rightarrow\mu$$X$) results
on proton and deuteron structure functions ;
a re-analysis of the CCFR ($\nu$Fe$\rightarrow lX$) data on $F_2$
and $xF_3$; new preliminary results from
CDF on charge asymmetry in $W$ production,
from E866 on Drell-Yan $\mu$-pair production and from
E706 on prompt photon production.
New results from HERA on $F_2$, on the gluon density at low $x$, 
on the charm contribution 
$F_2^{c\overline c}$, 
on a determination of $F_L$ 
and on measurements in the very low $Q^2$ region
are discussed.
}

\section{Introduction}

The study of the structure of hadrons has been always 
of great importance for the development of high energy physics.
This report concentrates on recent experimental
contributions to the understanding of the nucleon 
structure~\footnote{The spin structure of the nucleon, 
diffraction and hadronic jet production are discussed at this conference 
in the talks given by A. Bruell, E. Gallo and H. Schellman. 
Theoretical aspects are reviewed by S. Catani.}.   
Two main questions have been addressed in such studies.
Firstly, a test of the theory of strong interactions
and, secondly, a determination of the momentum distributions of the partons
within the nucleon.

The main sources of experimental information on the structure of the nucleon 
are the fixed target experiments with electron, muon, neutrino and 
proton beams, the $p\bar{p}$ colliders and the $ep$ collider at HERA.

Since the initial observation~\cite{scaling} of Bjorken scaling,
experiments on Deep Inelastic Scattering (DIS) play an outstanding role
in the investigation of the nucleon structure. 
This observation established that the quark-parton model
is a valid framework for the interpretation of data and that the DIS structure 
functions from different processes can be expressed in terms of 
universal parton densities. The later observation of
scaling violation~\cite{scaling-violation} 
and identification of partons as quarks
and gluons has 
confirmed the field theory of
quarks and gluons and their strong interactions, Quantum Chromodynamics (QCD). 
QCD in conjunction with
electroweak theory constitutes now the Standard Model of
elementary particle physics.

The differential cross section for neutral current deep inelastic
scattering of a charged lepton on a nucleon is related 
to the three structure functions $F_2$, $F_L$ and $xF_3$ according to
\begin{equation}                                                               
  \frac{d^2\sigma^{l^{\pm}N}}{dx dQ^2} 
  =\frac{2\pi\alpha^2}{Q^4x}
  \left[(1+(1-y)^2)F_2(x,Q^2)-y^2 F_L(x,Q^2)\mp (1-(1-y)^2)xF_3(x,Q^2)\right]
\label{dsigma}                                                                 
\end{equation}                                                                 
Here $Q^2$ is the squared four-momentum transfer
between the lepton $l$ and the nucleon $N$, 
$x$ denotes the Bjorken variable, and $y= Q^2/xs$ is the inelasticity,
where $s$ is the center of mass energy squared of the collision
and $\alpha$ is the fine structure constant.
In eq.~\ref{dsigma} it is assumed that $s$ is much larger 
than the nucleon mass.
For $Q^2$ much below the $Z^0$ mass squared,
the parity violating structure function $xF_3$ is negligible and
the structure function $F_2$ is given purely by photon exchange. 
The structure functions $F_2$ and $F_L$ are 
related by $R=\sigma_L/\sigma_T \simeq F_L/(F_2-F_L)$, where $R$
is the ratio of cross sections of longitudinally and transversely
polarized virtual photons. The contribution of the longitudinal structure 
function $F_L$ to the cross section
is important only at large $y$, typically $y \geq 0.4$.
In the quark-parton model $F_L$=0, and $F_2$ is a sum over the quark
and antiquark momentum fractions within a nucleon 
multiplied by the corresponding quark charge squared.

In perturbative QCD the structure function $F_2$ is a 
convolution of the parton distributions and coefficient functions $C(x,Q^2)$
\begin{equation}                                                               
  \frac{1}{x}F_2(x,Q^2) = 
     \sum^{n_f}_{i=1} e_i^2 C_i(x,Q^2)\otimes
     (q_i(x,Q^2)+\bar{q}_i(x,Q^2)) +
     C_g(x,Q^2)\otimes g(x,Q^2),
\label{f2-qcd}
\end{equation}
where $q_i$ denotes quarks of charge $e_i$ and $g$ gluons, $\otimes$ stands 
for a convolution integral and $n_f$ is the number of contributing flavors.
The parton distributions evolve with $Q^2$ following the DGLAP~\cite{DGLAP}
equations
\begin{equation}                                                               
  \frac{\partial}{\partial lnQ^2}\left(q\atop g\right) 
  = \frac{\alpha_s(Q^2)}{2\pi}
  \left[P_{qq} P_{qg}\atop P_{gq} P_{gg}\right]\otimes\left(q\atop g\right),
\label{evolution}
\end{equation}
where $\alpha_s$ denotes the strong coupling constant. 
The coefficient functions $C_i$ and splitting functions $P_{ij}$ 
are obtained by perturbative expansion
in a specific factorization and renormalization scheme.

The solution of the DGLAP evolution equation
to next-to-leading order (NLO) 
together with the input parton distributions, which cannot be derived
from first principles and should be fitted to data 
at some starting scale $Q_{\rm o}^2$, 
constitute the usual framework of a QCD analysis of the experimental results.

\begin{figure}[htbp]       \unitlength 1mm                               
\begin{picture}(100,90)
\put(22.,-4.){\psfig{file=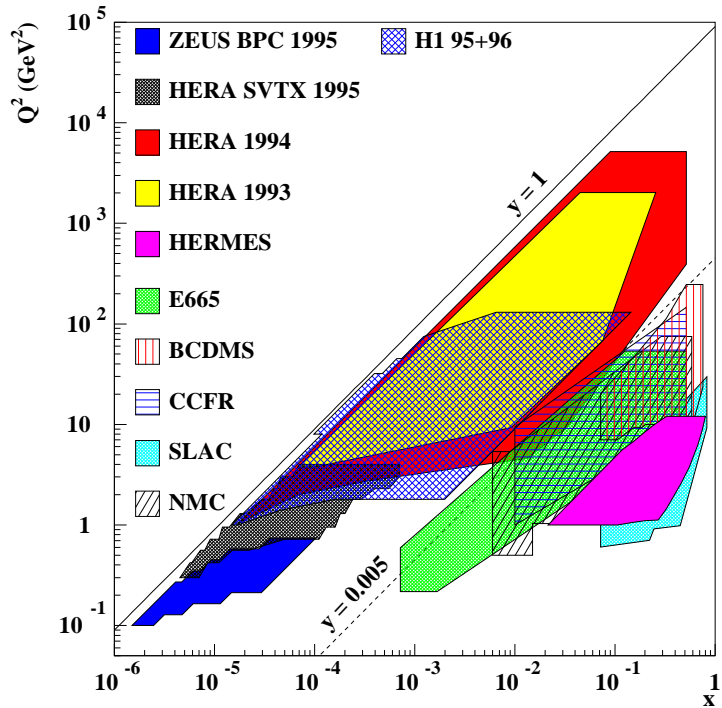,width=9.5cm}}
\end{picture}
\caption{The regions in $x$ and $Q^2$ covered by different DIS experiments.}
\label{xQ2plane}
\end{figure}                       

The kinematic phase space covered by DIS experiments
is summarized in Figure~\ref{xQ2plane}. The data span 6 orders of
magnitude in $x$ and 5 orders of magnitude in $Q^2$ and allow to
make very precise tests of QCD, extract quark and gluon densities 
and determine $\alpha_s$. 
Combined QCD analyses of the existing DIS information
together with information from hadron-hadron collisions is a
goal of NLO DGLAP fits such as performed by  
MRS~\cite{mrsdprim,MRSg,mrsR1,mrrs},
CTEQ~\cite{CTEQ,cteq4m,cteq-acot} and GRV~\cite{GRV}.

The paper is organized as follows.  
In section 2, devoted to results from fixed target DIS experiments,
the final NMC ($\mu$$N$) results
and a re-analysis of the CCFR ($\nu$Fe) data are presented.
Additional constraints on parton densities from hadron-hadron
collisions are presented in section 3. 
Recent HERA results on $F_2$, on the gluon density, on the charm contribution 
$F_2^{c\overline c}$ to the structure function of the proton, 
on the determination of $F_L$ by H1 
and on measurements in the very low $Q^2$ region
are discussed in section 4, followed by a  summary in section 5.

\section{DIS in Fixed Target Experiments}                  

Until a few years ago our knowledge of structure functions and
derived quantities such as parton distributions was almost entirely based 
on the fixed target experiments, using electron, muon and neutrino beams. 

\subsection{The Final NMC Results ($\mu$$N\rightarrow\mu$$X$)}

The New Muon Collaboration (NMC) at CERN has published final 
results~\cite{nmc-f2pd,nmc-f2pdratio,nmc-Adep,nmc-f2Sn}
on deep inelastic muon nucleon scattering at 
muon beam energies of 90, 120, 200, and 280 GeV.  

\begin{figure}[htbp]
\begin{picture}(100,230)(-15.,15.0)
\psfig{file=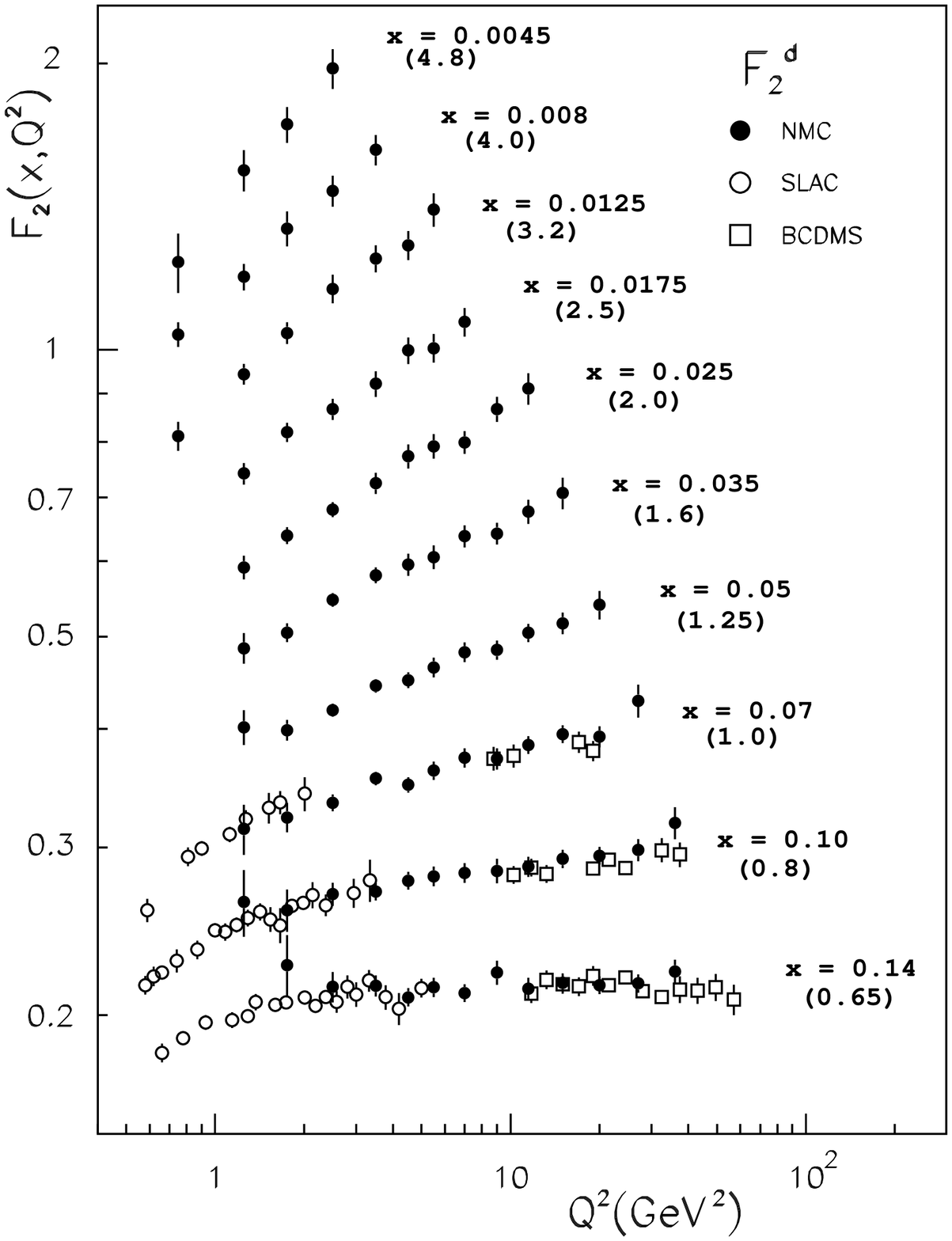,width=6.7cm}
\psfig{file=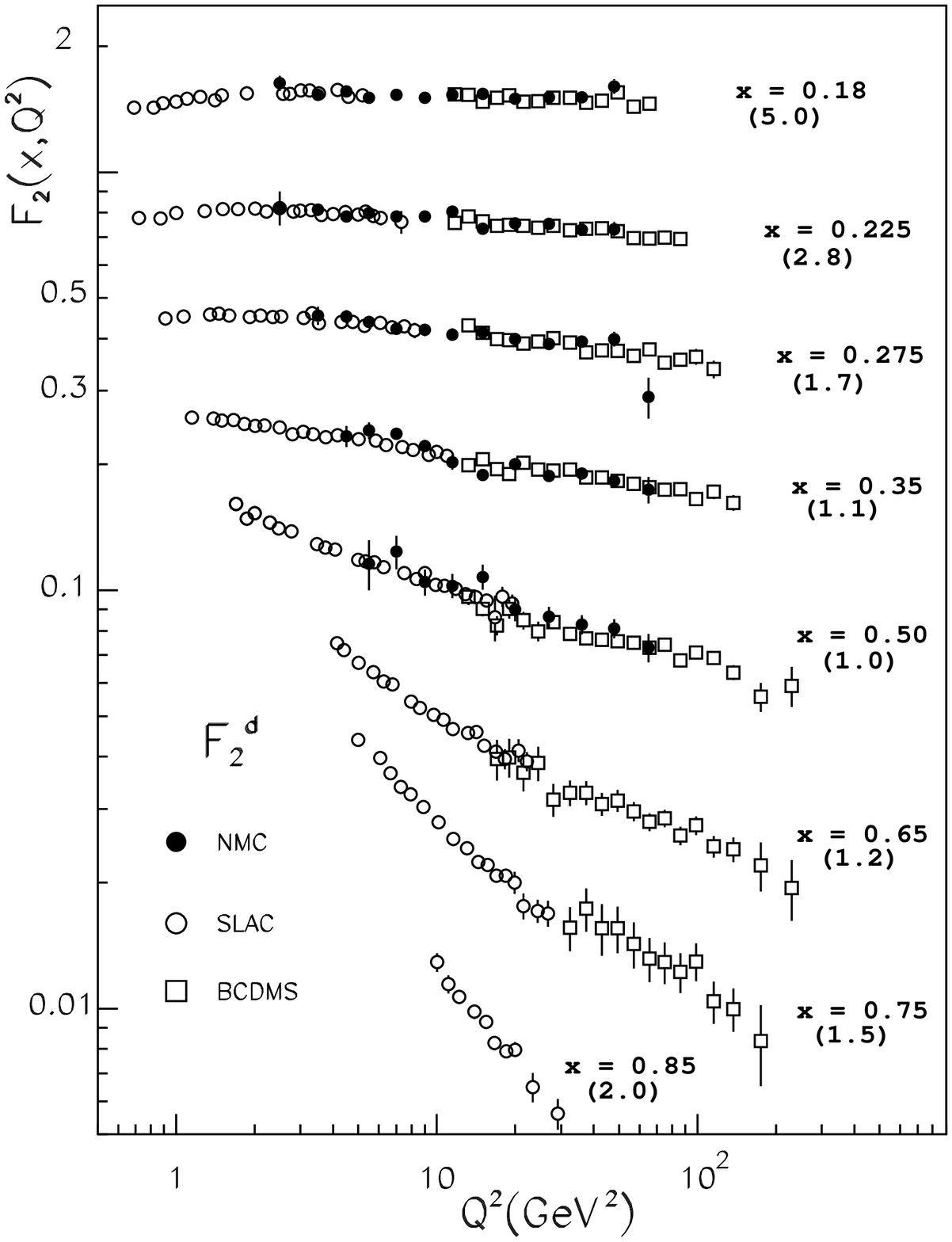,width=6.7cm}
\end{picture}                                                               
\caption{The final NMC results for $F_2^d$ compared to SLAC and BCDMS results.}
\label{f2d-nmc}
\end{figure}

The final NMC $F_2^p$ and $F_2^d$ data~\cite{nmc-f2pd} for
proton and deuteron targets 
(the results for $F_2^d$ are shown in Figure~\ref{f2d-nmc})
cover the kinematic range
$0.002 < x < 0.6$ and $0.5 <Q^2 < 75 $~GeV$^2$ with
high statistical accuracy and with systematic uncertainties 
between 1\% and 5\%. 
The coverage in $x$ and $Q^2$ was extended to lower values
due to the use of a small angle trigger.
The results compare well with previous measurements
from SLAC, BCDMS (see Figure~\ref{f2d-nmc}) 
and E665 and extrapolate smoothly to the
recent HERA data.
The ratio $R$ of cross sections of longitudinally and transversely
polarized virtual photons  was measured for
$0.002 < x < 0.12$ and $1 <Q^2 < 25 $~GeV$^2$. The results
are in agreement with earlier measurements as well as
with expectations from perturbative QCD.
The difference ($R^d - R^p$)~\cite{nmc-f2pdratio}, 
determined for $0.003 < x < 0.35$, is compatible with zero as expected.
 The results for the structure function ratio 
$F_2^d$/$F_2^p$~\cite{nmc-f2pdratio} 
cover the $x$ range from $0.001$ to $0.8$ and the $Q^2$ range
from $0.1$ to $145$ ~GeV$^2$ with a typical systematic accuracy of $0.5\%$.
The data on $F_2^d$/$F_2^p$ and the $F_2$ parameterization~\cite{NMC-old}
have been used to determine the Gottfried sum
$S_G=\int\limits_0^1(F_2^p-F_2^n)dx/x$. The result in the interval 
$0.004 < x < 0.8$ at $Q^2 = 4$~GeV$^2$ is 0.2281$\pm$0.0065
in agreement with the previous estimation~\cite{NMC-GSR}. 
The value obtained is below the expectation of 1/3, indicating
a flavor asymmetry in the quark-antiquark sea.

Nuclear effects were investigated studying the dependence on the
mass number $A$ in the shadowing region (small $x$), the enhancement
region (at $x$  about 0.1) and the EMC effect region (large $x$) 
by measuring with a series of different nuclei. A clear increase with $A$
was observed for all effects~\cite{nmc-Adep}.
A study of the $Q^2$ dependence of nuclear effects was performed 
using high luminosity measurements with thick carbon and tin targets
\cite{nmc-f2Sn}.
 
\subsection{Re-analysis of the CCFR Data ($\nu$Fe$\rightarrow lX$)}

\begin{figure}[htbp]
\begin{picture}(100,310)(-100.,7.)
\psfig{file=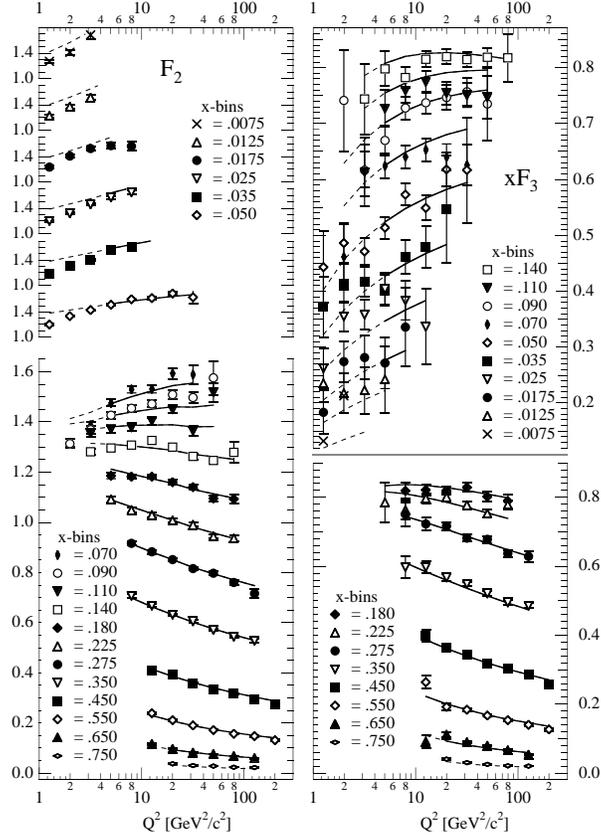,width=8.0cm}
\end{picture}                                                               
\caption{The updated $F_2$ and $xF_3$ data from CCFR.
The results of a NLO QCD fit are given by the solid line.
The dashed line extrapolates the fit to the lower $Q^2$ region
excluded from the fit.}
\label{ccfr-f2f3}
\end{figure}

Accurate $\nu$Fe structure function data have been available for some time
from the high statistics CCFR experiment at FNAL. 
In the earlier analysis~\cite{als-ccfrold}, 
the muon and hadron energy calibrations were
determined using a Monte Carlo technique. 
Recently the data have been re-analyzed~\cite{als-ccfr}
to determine $F_2$ and $xF_3$ 
for $0.0075 \leq x \leq 0.75$ and $1.3 \leq Q^2 \leq 126$~GeV$^2$
using the muon and hadron energy calibrations taken directly 
from test beam data.
The updated structure functions
corrected for radiative effects, for the non-isoscalarity of
the Fe target, for the charm-production threshold and for the mass of the
$W$-boson propagator 
are shown in Figure \ref{ccfr-f2f3}.

 The structure function $F_2$ from $\nu$Fe DIS can be compared
to $F_2$ from $e$ and $\mu$ DIS for an isoscalar target. 
To make this comparison, two corrections have been applied to the
charged-lepton data. The deuterium data from SLAC, NMC, and BCDMS 
have been corrected to Fe using the
$F_2^{lN}/F_2^{lD}$ ratio from SLAC and NMC. 
The second correction accounts for the electric charges of the quarks
participating in the electromagnetic interactions:
\begin{equation}                                                               
 \frac{F_2{^l}}{F_2{^\nu}} = 
\frac{5}{18}\left(1-\frac{3}{5}\frac{s+\bar{s}-c-\bar{c}}{q+\bar{q}}\right)
\label{f2lnu}                                                                 
\end{equation}                                                                 

\begin{figure}[htbp]
\vspace*{1.cm}
\begin{picture}(170,140)(-90.,7.)
\psfig{file=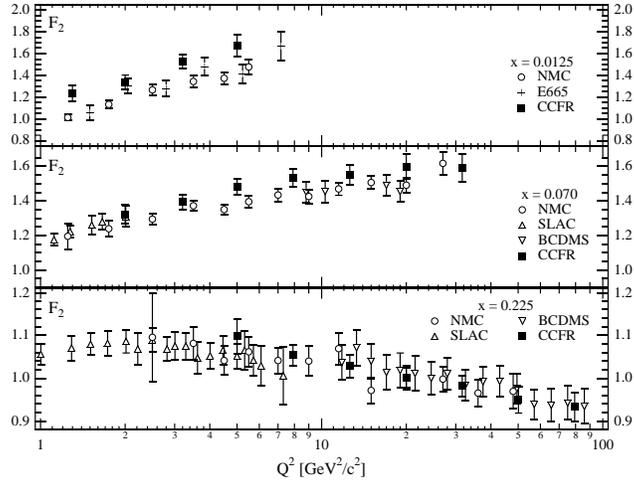,width=8.5cm}
\end{picture}                                                               
\caption{Comparison of the updated CCFR $F_2$ values for $\nu$Fe
with those for $\nu$D from NMC, E665, BCDMS and SLAC. 
The charged lepton data have been corrected to an isoscalar Fe target
and for quark-charge effects.}
\label{ccfr-nmc}
\end{figure}

The comparison of $F_2$ from the charged-lepton and neutrino DIS 
is shown in Figure \ref{ccfr-nmc}. The $F_2$ values generally agree well 
except in the low $x$ bin (0.0125), where there is  a 15\% discrepancy
between the NMC and CCFR results. It can not be explained by
increasing the size of the strange sea, as this is limited by CCFR
dimuon data~\cite{ccfr-dimuon}, 
however it has been suggested that its distribution may
be more complicated than usually assumed. Another possibility
is that the nuclear corrections are different for neutrino and 
charged leptons.

Using the improved $F_2$ and $xF_3$ data in the region
$Q^2 > 5$~GeV$^2$, $x < 0.7$ and $W^2 > 10$~GeV$^2$, 
the CCFR collaboration has
performed a QCD fit to extract $\Lambda_{QCD}$. Target mass corrections
were included into the fit. Higher twist (HT) effects were taken into
account. The best QCD fit to the data is shown in Figure \ref{ccfr-f2f3}.

From this fit in NLO QCD for 4 quark flavors
 the value $\Lambda_{\overline{MS}}$=337$\pm$28(exp.)$\pm$13(HT)~MeV 
has been obtained, which yields
$\alpha_S(M_Z^2)$=0.119$\pm$0.002(exp.)$\pm$0.001(HT)$\pm$0.004(scale)
\footnote{The NNLO analysis~\cite{kataev} of the new CCFR data
on $\nu$Fe$\rightarrow$$lX$ gives a very similar result: 
$\alpha_S(M_Z^2)$=0.117$\pm$0.002(exp.)$\pm$0.005(syst.)$\pm$0.003(theory).}.
A fit to the data on $xF_3$ only, which is not coupled to the gluon 
distribution, gives 
$\Lambda_{\overline{MS}}$ = 381$\pm$53(exp.)$\pm$17(HT) MeV,
which is consistent with the result of the combined fit of $F_2$ and $xF_3$
but has larger errors because effectively only half of the data are used.
The value of $\alpha_S$ is significantly higher than the earlier 
CCFR result~\cite{als-ccfrold},
$\alpha_S(M_Z^2)$=0.111$\pm$0.002(stat.)$\pm$0.003(syst.),
mainly due to the new energy calibration.
 
\subsection{Summary for DIS in Fixed Target Experiments}

The present fixed target program for unpolarized DIS with charged
lepton beams is now completed.
The final results of the SLAC, BCDMS, E665, NMC experiments
are published providing us with a firm basis for QCD analyses 
of the nucleon. 
The neutrino beam data from CCFR at FNAL
have been re-analyzed to give new $F_2$ and $xF_3$ values.
The value of $\alpha_s$ ($\alpha_s \sim 0.119$) extracted from 
the updated structure function results
is significantly higher than the earlier CCFR result ($\alpha_s \sim 0.111$).
It is also larger than the result~\cite{als-dis} based on
the SLAC/BCDMS data 
($\alpha_s \sim 0.113$)
and is very close
to the LEP values ($\alpha_s \sim 0.120$).
The existing very precise data sets 
are well consistent 
apart from a discrepancy of about 15\% at low $x$ between the $F_2$ values
derived from the CCFR and the NMC data.

\section{Constraints on Parton Densities from Hadron-Hadron Collisions}

Information on the valence quark density ratio $u(x)/d(x)$ at high
$Q^2$ and on the $\bar{u}(x)/\bar{d}(x)$ ratio of sea quarks
can be gained from $W$ production in $p\bar{p}$ collisions 
and  $\mu$-pair production via the Drell-Yan mechanism in $pp$
and $pd$ interactions. Prompt photon and jet data
from hadronic collisions are sensitive to the gluon density at large $x$. 
Recent preliminary results on these processes 
from CDF, E866 and E706 are presented in this section.

\subsection{Charge Asymmetry in W Production in $p\bar{p}$ Collisions}

In $p\bar{p}$ collisions, $W^+(W^-)$ bosons are produced primarily by
the annihilation of $u(d)$ quarks in the proton and $\bar{d}(\bar{u})$
quarks from the antiproton. As $u$ quarks carry on average more
momentum than $d$ quarks, the $W^+$'s tend to follow the direction
of the incoming proton and the $W^-$'s that of the antiproton. 
The charge asymmetry in the production of $W$'s as a function of rapidity 
is related to the $u$ and $d$ quark distributions at $Q^2\approx{M_W}^2$.
It is roughly proportional to the ratio of the difference and the sum
of the quantities $d(x_1)/u(x_1)$ and $d(x_2)/u(x_2)$, where $x_1$ and $x_2$
are the fractions of the nucleon momentum carried by the quarks in the $p$ 
and $\bar{p}$, respectively. 
Since the $W$ rapidity is experimentally
undetermined, 
\begin{figure}[htbp]
\begin{picture}(170,180)(-75.,7.)
\psfig{file=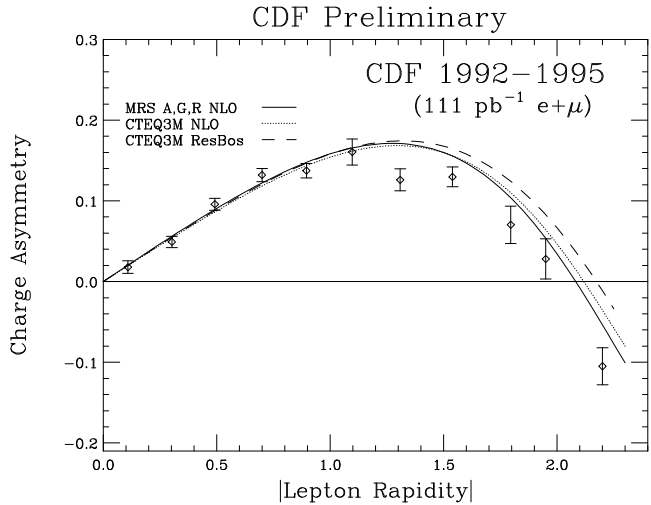,width=8.5cm}
\end{picture}                                                               
\caption{The charge asymmetry $A(y_l)$ corrected for detector
effects and backgrounds as a function of
the lepton rapidity $y_l$. 
Due to CP invariance $A(y_l)$ = -$A(-y_l)$
and the two values are combined.
The statistical and systematics errors are added in quadrature.}
\label{Wasym}
\end{figure}                                                                   
because of the unknown longitudinal momentum of the neutrino 
from  the $W$ decay, the lepton charge asymmetry is actually measured:
\begin{equation}                                                               
  A(y_l) =\frac{d\sigma^+/dy_l-d\sigma^-/dy_l}{d\sigma^+/dy_l+d\sigma^-/dy_l},
\label{Ayl}                                                                 
\end{equation}                                                                 
where $d\sigma^+(d\sigma^-)$ is the cross section for $W^+(W^-)$ 
decay leptons as function of the lepton rapidity~$y_l$.

Previously published $W$ asymmetry results obtained by the CDF collaboration
at FNAL \cite{CDF-Wasym-old} have been used already in 
global analyses~\cite{mrsdprim,CTEQ,GRV}
to extract parameterizations of
parton distribution functions in the nucleon. The new preliminary
CDF results~\cite{CDF-Wasym} are shown in Figure \ref{Wasym} and 
are based on the data from 1992 to 1995, corresponding to a five fold
increase in statistics. Furthermore the asymmetry measurement is extended
to larger rapidity (up to $|y_l|=2.2$) and provides information on 
parton densities in a larger $x$ range ($0.006 < x < 0.34$) 
than previously. As shown in Figure~\ref{Wasym},
the existing parameterizations are in good agreement with the new measurements
in the central region ($|y_l|<1.1$). However, at large rapidity 
expectations from global parton distribution analyses
are generally above the data. 

\subsection{Drell-Yan $\mu$-pair Production}                  

Dimuon production via the Drell-Yan mechanism can be used to 
investigate the question of flavor asymmetry in the nucleon sea.
The proton-nucleon Drell-Yan cross section can be written in
terms of parton distribution functions as
\begin{equation}                                                               
  \sigma^{pN} \sim \sum_{i} e_i^2 [q_i(x_1)\bar{q}_i(x_2)+q_i(x_2)\bar{q}_i(x_1)],
\label{sigpN}                                                                 
\end{equation}                                                                 
where $x_1$ and $x_2$ are the fractions of the nucleon momentum carried by 
the beam and target partons respectively. 
In the approximation $x_1 \gg x_2$ ,
the ratio of the Drell-Yan yields from protons incident on deuterium
and hydrogen targets, $\sigma^{pd}/{2\sigma^{pp}}$,
has a simple approximate relation to $\bar{u}/\bar{d}$:
\begin{equation}                                                               
{\frac{\sigma^{pd}}{2\sigma^{pp}}}|_{x_1 \gg x_2 } \approx 
\frac{1}{2} \left(1+\frac{\bar{d}(x_2)}{\bar{u}(x_2)}\right) 
\label{bardu}                                                                 
\end{equation}                                                                 

Preliminary results on the ratio of the deuterium to hydrogen 
\begin{figure}[htbp]
\begin{picture}(170,175)(-117.,-23.)
\psfig{file=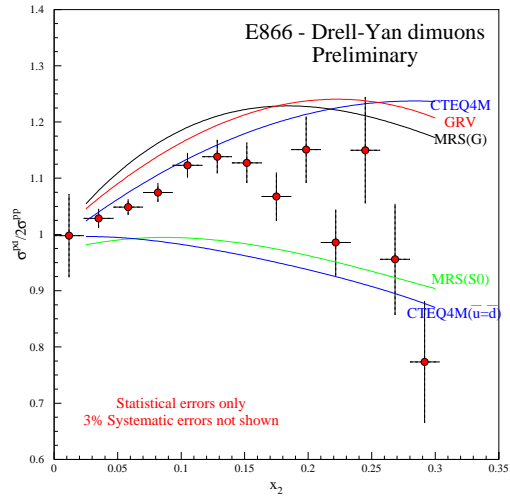,width=6.5cm}
\end{picture}                                                               
\caption{The ratio of the Drell-Yan cross sections
on deuterium and hydrogen targets.}
\label{DY}
\end{figure}                                                                   
Drell-Yan cross sections from the FNAL E866 experiment~\cite{E866} 
are shown in Figure \ref{DY} as a function of $x_2$.  
The data points
are compared with different parameterizations of the proton.
Also plotted is a curve based on CTEQ4M~\cite{cteq4m}, where
the parameterization was modified to force a flavor symmetric sea 
$\bar{u}_p = \bar{d}_p \equiv (\bar{u}_p+\bar{d}_p)/2 $ . 
The preliminary E866 results confirm the results of 
NMC~\cite{NMC-GSR,nmc-f2pdratio} and NA51~\cite{NA51}
that $\bar{d}_p>\bar{u}_p$.
The data in Figure~\ref{DY} are compatible with present parameterizations 
at low $x$, but for $x>0.2$ the parameterizations  fail. 
It is the final goal of the E866 experiment to measure the ratio
of Drell-Yan cross sections $\sigma^{pd}/{2\sigma^{pp}}$ 
with an accuracy of about 1\%
for $0.05 \leq x \leq 0.15$ 
and to determine $\bar{u}/\bar{d}$ over the full range up to $x \simeq 0.3$.

\subsection{Prompt Photon Production}                  

The prompt photon production $pN \rightarrow \gamma$$X$
is dominated by the subprocess $qg \rightarrow q\gamma$ and,
in leading order, directly related
to the gluon density. Recently the E706 experiment at Fermilab presented
high statistic measurements~\cite{E706} on large transverse momentum 
prompt photon and  inclusive $\pi^0$ cross sections 
using 530 and 800~GeV proton beams and a 515~GeV $\pi^-$ beam
incident on a Be target.
Current NLO QCD calculations failed to
describe the data, indicating the presence of a substantial
initial state parton transverse momentum ($k_T$) in the hard scattering
(a discussion on the $k_T$ problem can be found in ref.~\cite{Huston}).
A simple implementation of a parton $k_T$ in QCD calculations,
using empirical $<k_T>$ values consistent with observations, 
provides a reasonable
description of the data. The gluon distribution obtained in the
combined fit (taking into account $k_T$ effects) 
to the DIS, Drell-Yan and E706 prompt photon data 
is similar to the CTEQ4 result~\cite{cteq4m} and consistent with 
the jet cross section results from CDF and D0.
An improved theoretical understanding of soft gluon
effects will facilitate the determination of the gluon distribution
function at high $x$.


\begin{figure}[htbp]
\begin{picture}(170,270)(0.,-60.)
\put(3.,0.){\psfig{file=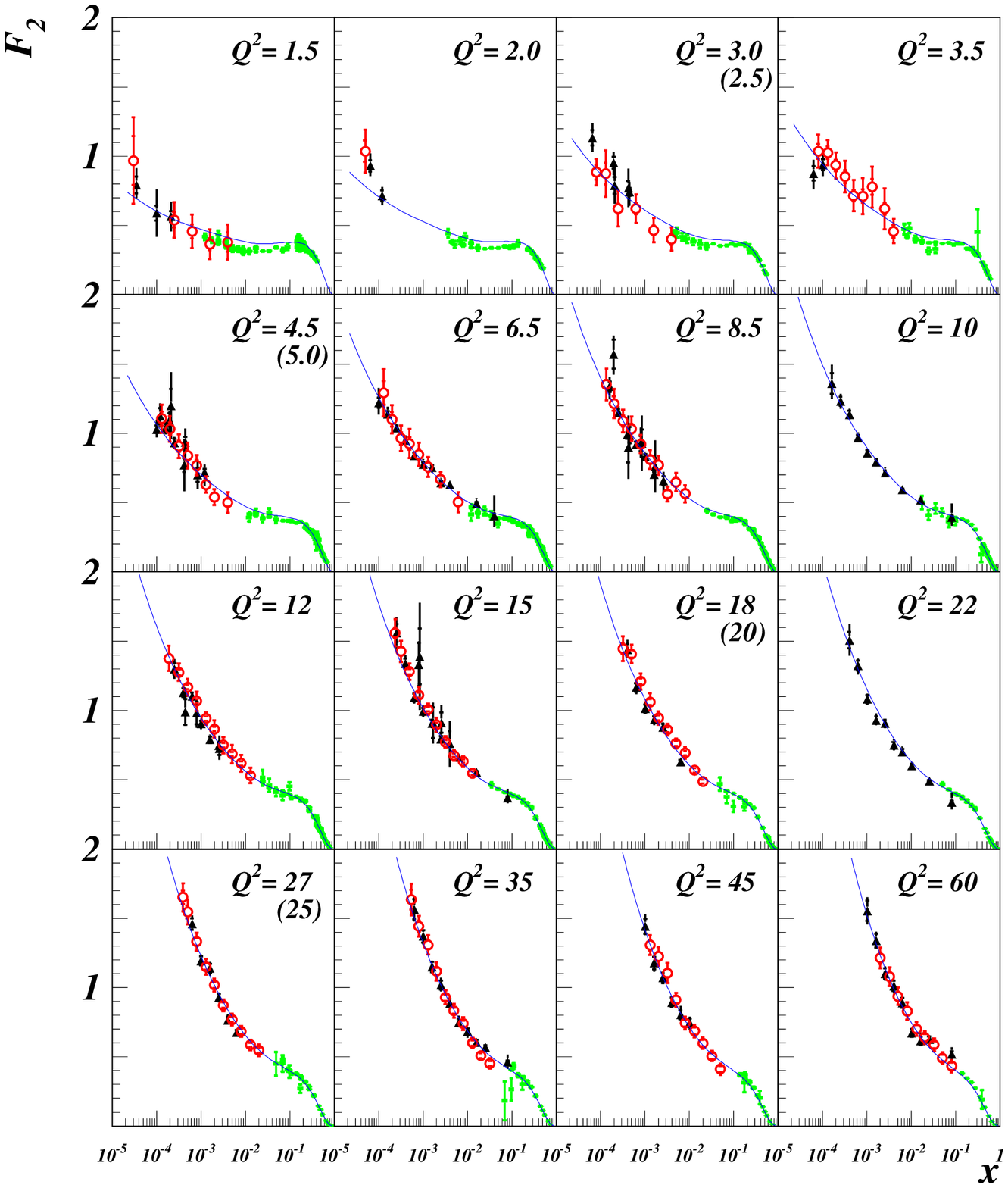,width=8.2cm}}
\put(220.,0.){\psfig{file=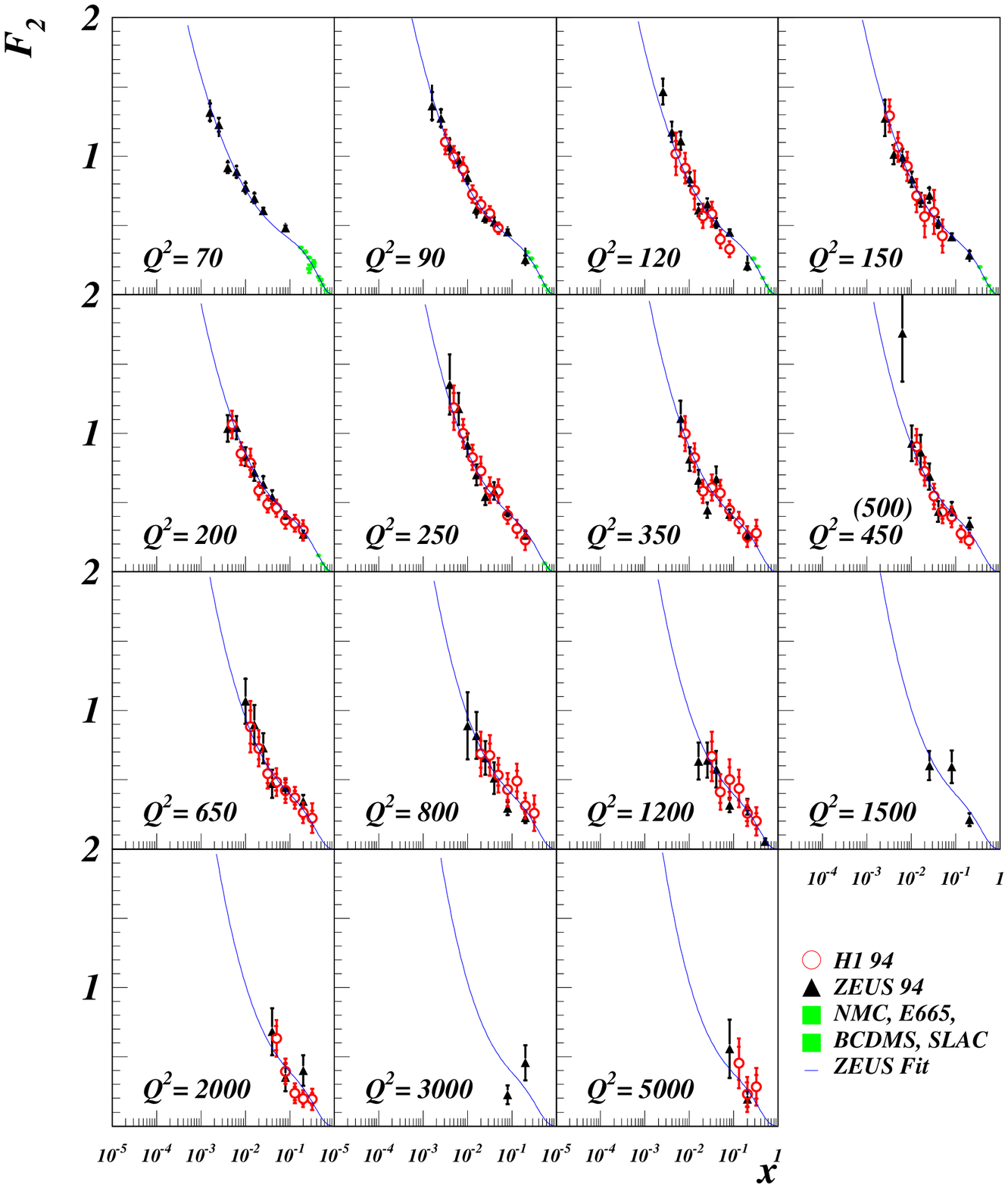,width=8.2cm}}
\end{picture}                                                               
\caption{$F_2$ data from HERA (1994) and fixed target experiments 
         at fixed $Q^2$ (in GeV$^2$) as a function of $x$. 
         The lines correspond to the NLO QCD fit by ZEUS.}
\label{f2-94all}
\end{figure}         

\section{The HERA Results}

Experiments at HERA extended the previously accessible kinematic range
up to very large values of $Q^2 > 10^3$~GeV$^2$, 
and down to very small values of
$x < 10^{-4}$ (Figure~\ref{xQ2plane}).
The first $F_2$ measurements reported at 
HERA~\cite{f2-h192,f2-zeus92}, based on data collected in
1992, revealed a pronounced rise of $F_2$ 
at low $x < 10^{-2}$ with decreasing $x$.
The rise was confirmed by the much improved data of 
1993~\cite{f2-h193,f2-zeus93}.
This rise can be understood as an increase in the quark-antiquark sea 
which in turn is being driven (eqs.~\ref{f2-qcd},~\ref{evolution})
by a rapid increase in the gluon density.
Thus, the quantitative investigation of gluon 
dynamics at low $x$ is one of the major challenges at HERA.

The first substantial data samples, 
with an integrated luminosity of about 3 pb$^{-1}$,
have been collected in 1994.
Using this data, the two HERA experiments, H1 and ZEUS,  
have published $F_2$ results~\cite{f2-h194,f2-zeus94,fl-h194} 
covering a range in $Q^2$, $x$, and $y$, 
corresponding to $1.5 <Q^2 < 5000 $~GeV$^2$, 
$3\cdot10^{-5} < x < 0.5$ and roughly $0.01<y<0.6$.
In 1995, both ZEUS and H1
have improved their detectors to be able to measure 
electron scattering angles close to zero.
ZEUS added a special detector (BPC)
near the beam pipe in the electron beam (backward)
direction which allowed to measure $F_2$ down to 
$Q^2$=0.11~GeV$^2$ \cite{f2-bpc}.
\begin{figure}[htb]
\begin{picture}(170,355)(-85.,-98.)
\psfig{file=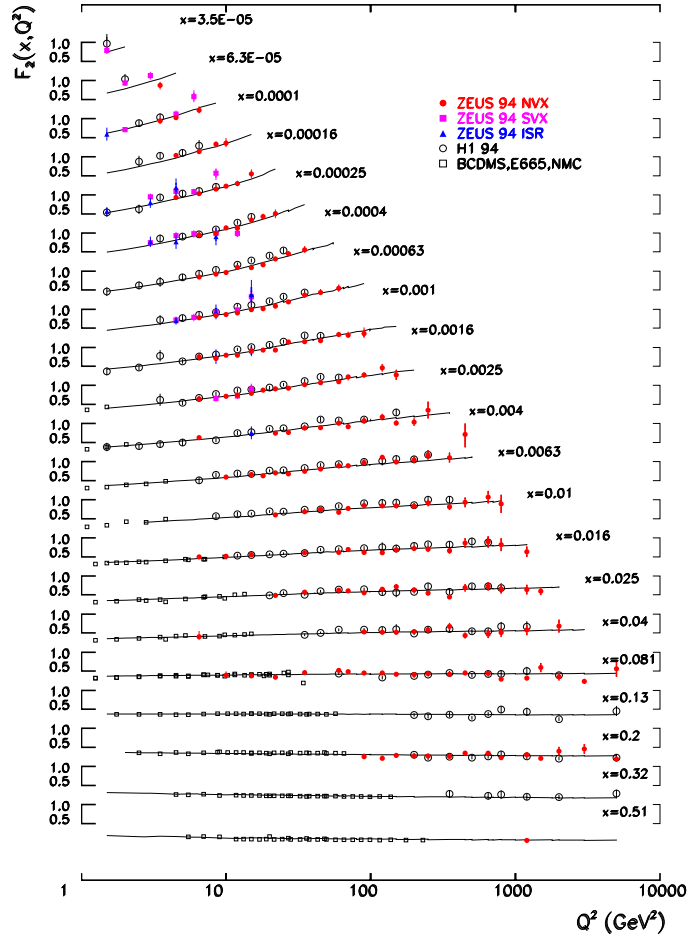,width=9.5cm}
\end{picture}                                                               
\caption{$F_2$ data from HERA (1994) and fixed target experiments 
         at fixed $x$ as a function of $Q^2$.
         The lines correspond to the NLO QCD fit by ZEUS.}
\label{f2xq2-94}
\end{figure}         
The H1 collaboration replaced the previous electromagnetic calorimeter 
in the backward direction by 
a lead/scintillating fiber calorimeter (SPACAL)~\cite{spacal} and measured 
$F_2$ down to $Q^2$=0.35~GeV$^2$~\cite{f2-h195} 
using data collected during a short period in 1995
when the $ep$ collision
vertex was shifted by 70 cm in the proton-beam direction 
with respect to the nominal position.
With this new calorimeter, using 1996 data, H1 measured the cross section 
up to $y$=0.82, where the sensitivity
to the longitudinal proton structure function $F_L$ is 
enlarged (eq.~\ref{dsigma}).
The region of very large $Q^2 > 15000$~GeV${^2}$, 
although limited by event statistics,
became recently of high interest and is discussed 
elsewhere~\cite{Straub,Elsen}.

All existing $F_2$ data from HERA were analyzed in the
framework of perturbative QCD with the goal to determine 
the gluon distribution.
A quantity directly related to the gluon density is
the charm contribution $F_2^{c\overline c}$ to the structure 
function of the proton at low $x$. 
It was measured by both collaborations.
Another quantity related in QCD to the gluon~\cite{fl-qcd}
is the longitudinal proton structure function $F_L$.
The H1 collaboration made an attempt to 
derive $F_L$ from the measured cross section at high $y$ 
assuming that $F_2$ is given by a QCD fit to data at lower $y$.
The transition between the region of perturbative QCD (DIS)
and Regge phenomenology (photoproduction, $Q^2 = 0$) was studied
using the HERA measurements in the very low $Q^2$ region. 

\begin{figure}[htb]
\begin{picture}(170,305)(-90.,-80.)
\psfig{file=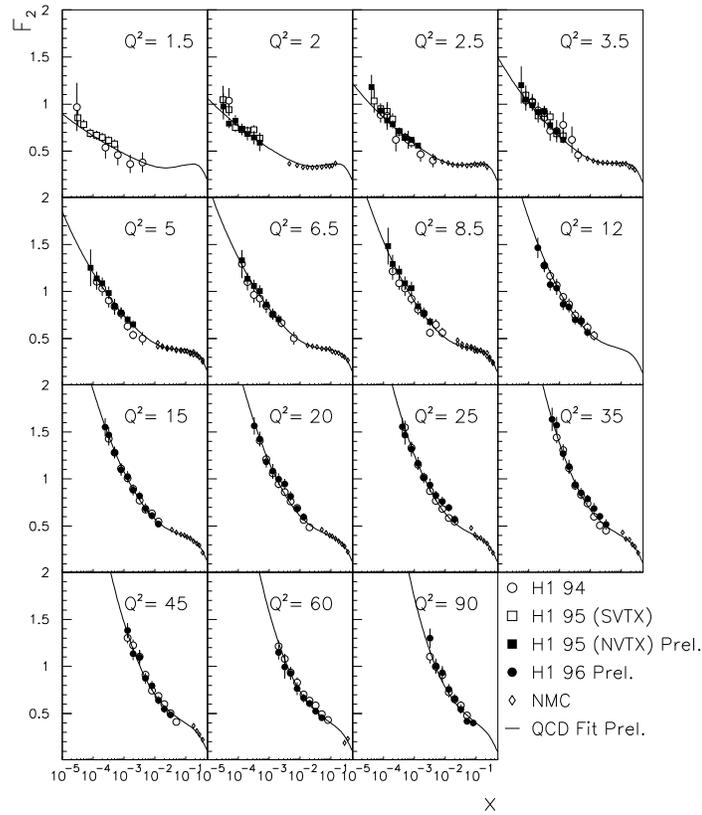,width=9.5cm}
\end{picture}                                                               
\caption{Measurements of $F_{2}$ by the H1 experiment. The new data
(1995 data for $2 < Q^{2} < 8.5$ GeV$^{2}$ and
1996 data for $12 < Q^{2} < 90$ GeV$^{2}$) are in good 
agreement with the previous results on $F_{2}$.
The curves represent the preliminary result of a NLO QCD  fit 
to the H1, NMC and BCDMS structure function data, described in Section 4.2.} 
\label{f2-9596}
\end{figure}         

\subsection{The Proton Structure Function 
$F_2(x,Q^2)$ at HERA}                                       
                                                          
The HERA results for the structure function $F_2$ from the 1994 data
are shown in Figure \ref{f2-94all} as function of $x$
and in Figure~\ref{f2xq2-94} as function of $Q^2$.
$F_2$ was derived from the $ep$ cross section according to eq.~\ref{dsigma}.
The values of $R$ needed for that were calculated using 
the QCD relation~\cite{fl-qcd}
and the result of a NLO QCD fit (ZEUS)
or the GRV parameterization~\cite{GRV} (H1).
The typical systematic error is around 5\% and dominates the total error
everywhere apart from the high $Q^2$ region. 
The data from H1 and ZEUS are consistent with each other 
and smoothly connected to data from the fixed target experiments.   
The steep rise of $F_2$ with decreasing $x$ and the scaling violation 
are clearly visible in Figure~\ref{f2-94all} 
and in Figure~\ref{f2xq2-94} respectively.
The curves in Figures~\ref{f2-94all},~\ref{f2xq2-94} represent results
of the NLO QCD fit by ZEUS.

Figure \ref{f2-9596} shows new, preliminary H1 results~\cite{f2-h196}
on $F_2$ from the data taken in 1995 and 
1996 with $ep$ interactions at the standard (nominal vertex) point.
The previous measurements of H1 
(from 1994 and shifted vertex running in 1995) and 
the higher $x$ data of NMC are given as well.
There is remarkable agreement with the 1994 data although 
those were taken with a different apparatus in the backward direction. 
The curves in Figure~\ref{f2-9596} represent a NLO QCD fit by H1
which is used for a determination of the gluon density as described below. 

\begin{figure}[htb]
\begin{picture}(170,225)(-90.,8.)
\psfig{file=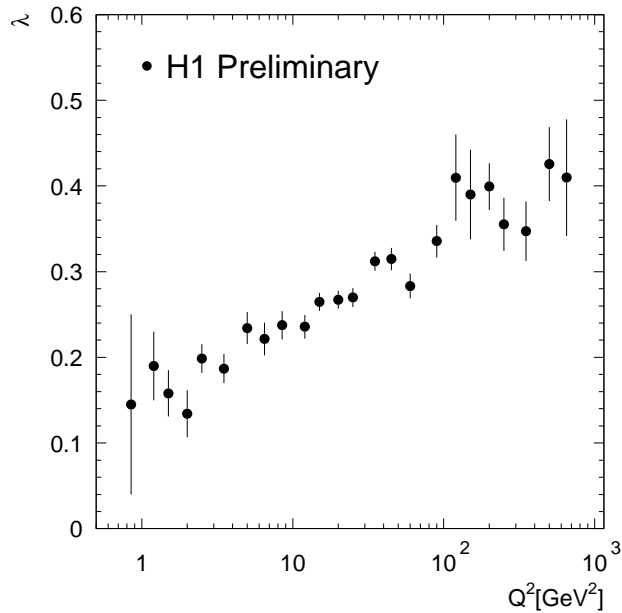,width=9.0cm}
\end{picture}                                                               
\caption{Variation of the exponent $\lambda$ from fits 
to the H1 data (1994, 1995 and 1996) of the form
$F_2 \propto x^{-\lambda}$ at fixed $Q^2$ values and $x < 0.1$.}
\label{lambda-9596}
\end{figure}

The rise of $F_{2}$ towards low $x$
has been quantified by determining the 
exponent $\lambda$ of $F_2 \propto x^{-\lambda}$ at fixed $Q^2$
(or equivalently $F_2 \propto W^{2\lambda}$, where $W \approx \sqrt{Q^2/x}$
is the center of mass of the $\gamma^*p$ system). 
Figure \ref{lambda-9596} 
represents a preliminary update of the previous H1 result
\cite{f2-h194,f2-h195} on the exponent $\lambda$.
There is a smooth 
transition visible from large values of $\lambda \simeq 0.40$ for 
$Q^{2} = 1000$ GeV$^{2}$ down to $\lambda \simeq 0.15$ for
$Q^{2} = 1$ GeV$^{2}$ approaching $\lambda \simeq 0.08$
which has been measured for hadronic and real photoproduction ($Q^{2} = 0$)  
total cross sections.
The results from HERA for the transition region between 
DIS and photoproduction
are discussed in detail in section 4.5.

\subsection{The Gluon Distribution $xg(x,Q^2)$ at Low $x$}


The scaling violation, which is clearly visible in Figure~\ref{f2xq2-94}
for the structure function $F_2$ in the HERA domain at low $x$,
is caused by gluon bremsstrahlung from quarks and quark pair production
from gluons and is related to the gluon density.
Both the H1 and the ZEUS collaborations performed NLO QCD fits
to their $F_2$ data with the goal to determine the gluon distribution 
at low $x$. 
The fits use
the $\overline{MS}$ renormalization scheme with 
the DGLAP evolution equations \cite{DGLAP}
for three light flavors adding the charm
contribution determined in the NLO calculation of the boson gluon 
fusion (BGF) process \cite{lae2}.

The ZEUS fit was performed to ZEUS data~\cite{f2-zeus94} 
in the range $1.5 \leq Q^2 \leq 5000$~GeV${^2}$.
The input scale  $Q_{\rm o}^2$ was chosen to be 7~GeV${^2}$, at which the
gluon distribution $xg$, the singlet quark distribution $x\Sigma$ and
the difference of up and down quarks in the proton $x\Delta_{ud}$
were parameterized as
\begin{eqnarray}\label{input-zeus}
xg(x)&=& A_gx^{\delta_g}(1-x)^{\eta_g}(1+\gamma_{g}x),\nonumber \\
x\Sigma(x)&=& A_{s}x^{\delta_{s}}(1-x)^{\eta_{s}}
(1+\varepsilon_{s}\sqrt{x}+\gamma_{s}x),\nonumber \\
x\Delta_{ud}(x)&=& A_{ns}x^{\delta_{ns}}(1-x)^{\eta_{ns}}.
\end{eqnarray}
The strange quark distribution was assumed to be 20\% of the sea
at $Q^2 = 4$~GeV${^2}$~\cite{ssea}. The sea quark density is obtained 
by subtracting the valence distribution (taken from the MRSD-$^{'}$
parameterization~\cite{mrsdprim}) from the singlet distribution.

In the present update of the published H1 fit results~\cite{f2-h194} 
the starting point of the evolution was chosen 
to be $Q_{\rm o}^2 = 1$ GeV$^2$ and all H1 data 
with $1.5 \leq Q^2 \leq 5000$~GeV${^2}$ 
including the measurements presented at this conference
(see previous section)  were included in the fit. 
In order to reduce the influence of the longitudinal structure 
function a cut of $y < 0.6$ was used for all H1 data sets.
The input parton distributions at the 
starting scale $Q^{2}_{\rm o}$ were parameterized as follows:

\begin{eqnarray}\label{input-h1}
xg(x)&=& A_gx^{B_g}(1-x)^{C_g},\nonumber \\
xu_v(x)&=& 
A_{u}x^{B_{u}}(1-x)^{C_{u}}(1+D_{u}x+E_{u}\sqrt{x}),\nonumber \\
xd_v(x)&=& A_{d} x^{B_{d} 
}(1-x)^{C_{d}}(1+D_{d}x+E_{d}\sqrt{x}),\nonumber \\
xS(x)&=& A_{S} x^{B_{S} }(1-x)^{C_{S}}(1+D_{S}x+E_{S}\sqrt{x}),
\end{eqnarray}
where $S=\bar{u}+\bar{d}$ and $\bar{u}=\bar{d}= 2\bar{s}$ define the sea 
distributions.

\begin{figure}[hbt]
\begin{center}
\begin{picture}(200,230)
\put(-30,15){\psfig{file=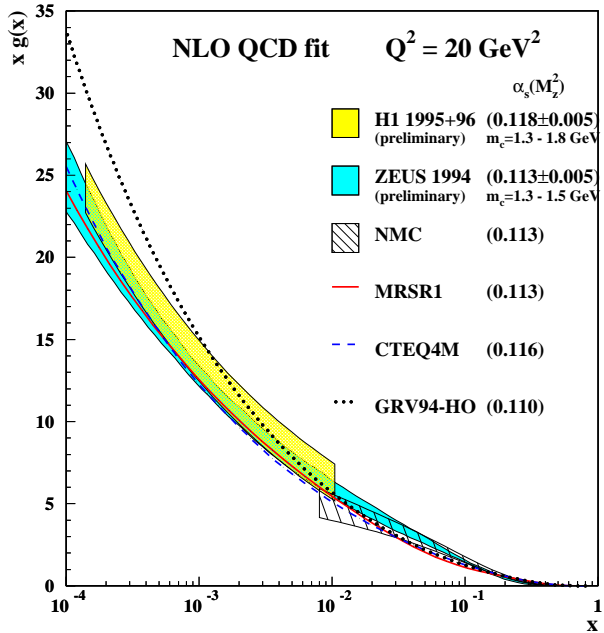,width=10.0cm}}
\end{picture}
\end{center}
\caption{Gluon distributions from the HERA NLO QCD fits to  
   structure function data. The error bands include the statistical 
   and systematic errors and also the uncertainties due to $\alpha_{s}$, 
   due to the charm quark mass $m_{c}$ and due to 
   the loosely constrained behaviour of $xg$ at high $x > 0.1$ (only H1 fit).
   The NMC fit is also shown at higher $x$. 
   The MRSR1, CTEQ4M, and GRV94-HO parameterizations are shown for
   comparison.}
\label{xg}
\end{figure}
 
In order to constrain the valence quark densities at higher $x$, 
proton and deuterium data of the muon scattering
experiments NMC~\cite{nmc-f2pd} (H1 and ZEUS fits) 
and BCDMS~\cite{BCDMS} (H1 fit only) were also used.
To avoid a possible  influence of higher twist effects, data in the range  
$ x > 0.5$ for $Q^2<15~$GeV$^2$ were excluded from the H1 fit.
The normalizations of all data sets were allowed to vary 
taking into account the quoted errors.

The resulting gluon distributions as determined by the HERA 
experiments~\cite{gluon-zeus94,f2-h196}
are shown in Figure \ref{xg}  for $Q^2 = 20$~GeV$^2$. 
The error bands account for statistical and systematical errors including
correlations. They take also into account possible variations 
of $\alpha_s$ by $\pm$0.005 around $\alpha_s$=0.113~\cite{als-dis}
(0.118) in case of ZEUS (H1) and of the charm mass 
$1.3 < m_c < 1.5$ GeV for ZEUS and $m_c$ = 1.5$\pm$0.3 for H1.
The agreement between the fit results is good. 
At the lowest $x$ values, $x\sim 10^{-4}$, 
the gluon distribution is now determined with a precision
of about 10\%. At $x \approx 0.01$ the HERA fits make contact with the fit 
performed by the NMC collaboration on their own data~\cite{xg-NMC}.

The gluon distribution in Figure \ref{xg} is rising steeply towards low $x$. 
For comparison the gluon densities from the
recent parton distributions MRSR1~\cite{mrsR1}, 
CTEQ4M~\cite{cteq4m} and from GRV94-HO~\cite{GRV}
are shown in the Figure as well.
Whereas the agreement with MRS and CTEQ at low x is good, 
the gluon obtained from the dynamical evolution by GRV
is too steep.

In global analyses, non-DIS measurements like prompt photon
and/or jet data are generally used 
to constrain the very high $x$ region (see section 3.3).
These data sets have not been included in the HERA fits.
This was taken into account by the H1 collaboration 
as an additional uncertainty 
which was estimated by 
a control fit with a five parameter gluon distribution forced 
to reproduce the high $x$ gluon density of ref.~\cite{MRSg}. 
This leads to
a gluon distribution which is  lower by  nearly 10\% at $x=0.01$ but
in very good agreement with the standard three parameter
gluon at lower $x$. The  difference of
these two determinations has been included in the error band 
in Figure \ref{xg} and is a dominant contribution to 
the error of $xg$ at $x$ near to 0.01.

\subsection{Charm Contribution 
{$F^{c\overline c}_2(x,Q^2)\;$} to the Proton
Structure Function}

The H1 and ZEUS experiments have published results on open charm production 
in deep inelastic scattering~\cite{f2c-h1,f2c-zeus} 
based on the 1994 data.
Recently the ZEUS collaboration presented an update~\cite{f2c-zeus95} 
of their results
using the 1995 data, doubling the statistics and
widening the $Q^2$ coverage.

Tagging of charm events is performed by reconstructing 
$D^{*+}$ \footnote{Charge conjugates are always implied.} 
and $D^0$\footnote{This channel is used only in the H1 experiment.} 
mesons via their decays into 
$D^{*+}\rightarrow D^0\pi^+_s\rightarrow(K^-\pi^+)\pi^+_s$
and $D^0\rightarrow K^-\pi^+$, respectively.
For the $D^{*+}$ 
analysis the mass difference $\Delta m= m(D^{*+})-m(D^0)$, being very close to
the $\pi$ mass, leads to a good resolution and signal to background 
ratio.
A satisfactory suppression of the combinatorial background is obtained 
also in the $D^0$ analysis by making
use of the hard fragmentation of charm quarks. 

\begin{figure}[hbt]    
    \begin{picture}(170,270)(-85,20)
    \put(0,0){
 \includegraphics{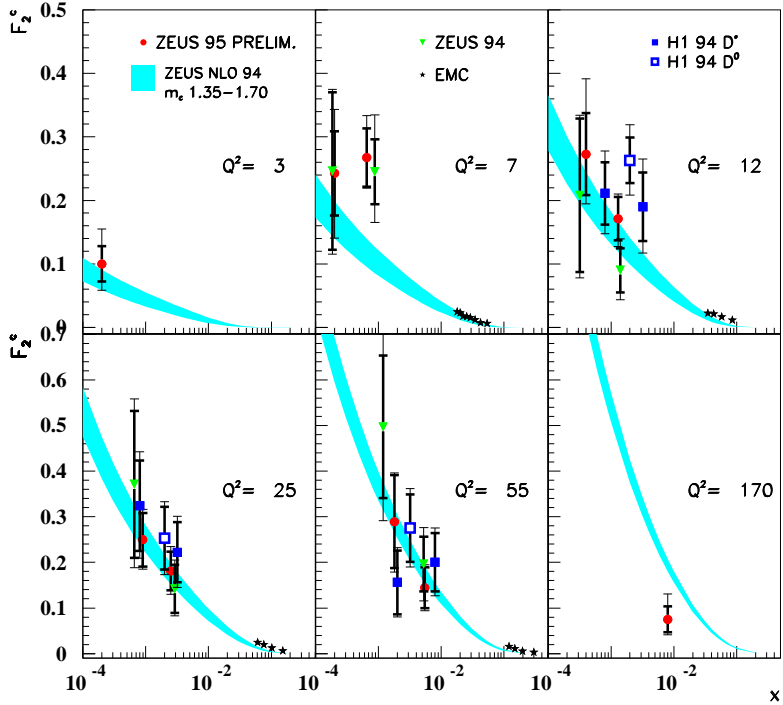}}
\end{picture}
\caption{
    The $F_2^{c\overline c}$ results from H1, ZEUS, and EMC are
    shown as function of $x$ for different bins of $Q^2$ (in GeV$^2$).
    The shaded band represents the NLO calculations 
    with different charm masses based on BGF using 
    the gluon density extracted from the ZEUS NLO DGLAP fit to 
    inclusive $F_2$.}  
\label{f2c}
\end{figure}

The charm contribution $F^{c\overline c}_2(x,Q^2)$ to the structure function
is obtained by applying the relation to the  
one photon exchange cross section for charm production
\begin{equation}
\displaystyle
\frac{d^2\sigma^{c\overline c}}{dxdQ^2}=\frac{2\pi\alpha^2}{Q^4x}
\left(1+\left(1-y\right)^2\right)\;F^{c\overline c}_2(x,Q^2)\;,
\end{equation} 
with the assumption that $R = 0$.  
$\sigma^{c\overline c}$ is obtained 
from the $D^{*+}$ and $D^0$ cross sections 
by integration and extrapolation
outside the measured range 
in transverse momenta and pseudo-rapidities of the $D^{*+}$, $D^0$ mesons
using NLO calculations~\cite{cc-NLO}. The calculations are 
based on the boson gluon fusion (BGF) production mechanism
and the gluon distribution obtained
by NLO DGLAP QCD fits to the inclusive $F_2$ data (see previous section). 

Figure \ref{f2c} shows $F^{c\overline c}_2(x,Q^2)$ as measured by H1 
and ZEUS together with the EMC results~\cite{emc}. 
The HERA measurements extend our knowledge of 
$F^{c\overline c}_2(x,Q^2)$ by two orders of magnitude
towards smaller $x$ values.
The charm contribution, $F^{c\overline c}_2(x,Q^2)$, to the proton
structure function
is seen to rise by about one order of magnitude 
from the high $x$ region covered by the fixed target experiment 
to the low $x$ region measured at HERA.
Averaged over the kinematic range at HERA, the ratio 
$F_2^{c\overline c}/F_2$ is about 25\%.

In Figure \ref{f2c} the data are also
compared with NLO QCD calculations for $F^{c\overline c}_2(x,Q^2)$
shown as a band, where the upper and lower limit corresponds to
a charm quark mass of 1.35 and 1.7 GeV, respectively.
The measured rise of $F^{c\overline c}_2(x,Q^2)$ from the
high to the low $x$ region is reasonably described 
in the three flavor number scheme with charm production via boson gluon fusion.
More precise data are needed to study the details of
the charm production mechanism and to
distinguish between different approaches,
MRRS~\cite{mrrs}, CTEQ~\cite{cteq-acot} and BMSN~\cite{buza}, which
provide a consistent treatment of heavy quark production from the
threshold region $Q^2\approx m_q^2$ to the asymptotic region
$Q^2\gg m_q^2$.

The first results from HERA on $F^{c\overline c}_2(x,Q^2)$ are very
promising. They indicate that 
the BGF process is dominant. However, high precision
results are still to come with high luminosity at HERA and the use of
silicon micro-vertex detectors, installed already by H1 and
planned by ZEUS for the year 2000, which could improve the
detection efficiency by an order of magnitude.

\subsection{The Longitudinal Proton Structure Function $F_L(x,Q^2)$}
 The H1 measurements \cite{f2-h194,fl-h194,f2-h196} of $F_2$, shown
in Figure~\ref{f2-9596} and discussed in section 4.1, 
were limited to $y$ values below 0.7, and 
the contribution of $F_L$ to the cross section was estimated by
a QCD calculation.
At large $y$ the  weights of $F_2$ and $F_L$ in eq.~\ref{dsigma}
become of comparable size.
The H1 collaboration has attempted to reverse that procedure, 
i.e. to measure the cross section at 
the maximum possible $y$ and to derive $F_L$ assuming that $F_2$ 
can be obtained from a QCD fit to data at lower $y$. 

The measured DIS cross sections for $Q^{2}$ between 12 
\begin{figure}[hbt]
\begin{center}
\begin{picture}(170,360)
\put(0,-25){
\includegraphics{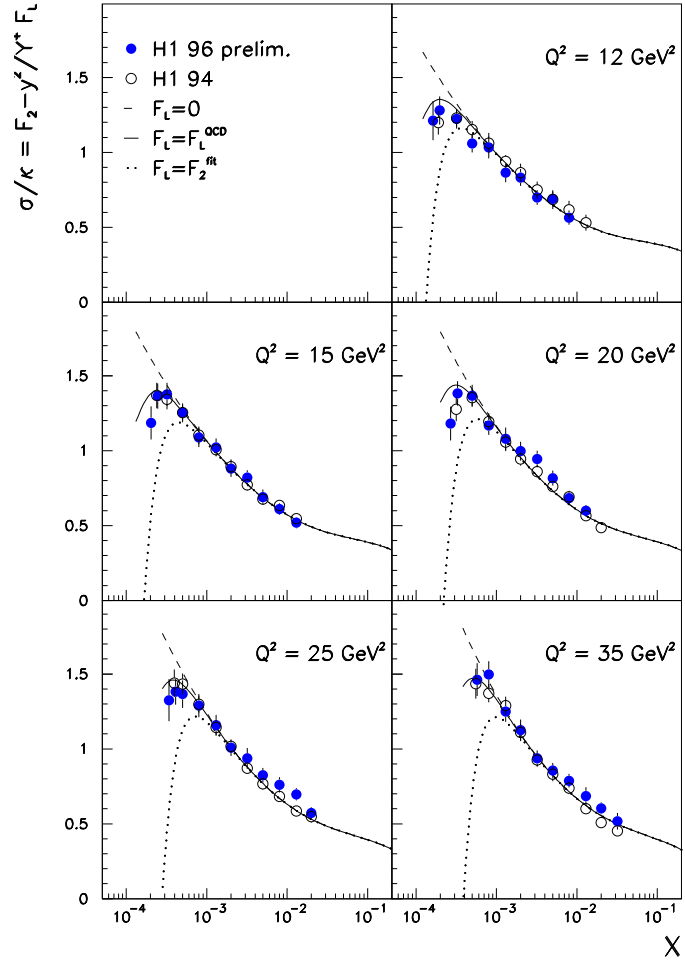}}
\end{picture}
\end{center}
   \caption{Measurement of the DIS cross section by H1 (1994, 1996)
   divided by the kinematic factor $\kappa=(2 \pi \alpha^{2} 
   \cdot Y_{+})/(Q^{4}x)$, where $Y_{+} = 1 + (1-y)^2$.
   The solid line shows the QCD calculation of $F_2$ and of $F_L$.
   Also shown are the cross section calculations
   using the  extreme assumptions $F_{L}=F_2$ (dotted line)
   and $F_L=0$ (dashed line). The lowest $x$ points for 
   $12  \leq Q^{2} \leq 25$~GeV$^2$ correspond to $y = 0.82$. }
\label{sig}
\end{figure}
and 35 GeV$^{2}$ are shown in Figure \ref{sig} comparing the 1994 data 
(open points) with the preliminary data of 1996 (closed points). 
The error bars include statistical and systematic errors added in quadrature.
Both cross section measurements agree well. 
The new data extended the $y$ range at four $Q^{2}$ values to $y=0.82$, 
thus considerably increasing the sensitivity to $F_{L}$. 
The total systematic errors of the cross section at the largest $y$ is 8\%, 
rather independently of $Q^{2}$.
 
Figure \ref{sig} shows also calculations of the cross 
section using the  QCD fit to $F_{2}$, described in
section~4.2, and three different 
assumptions on the longitudinal structure function $F_{L}$. The 
measured cross section is in  agreement with the NLO DGLAP 
calculation apart from large $y$ at $12  \leq Q^{2} \leq 25$~GeV$^{2}$, 
where the measured points tend to be lower than the QCD 
expectations (solid lines in Figure~\ref{sig}).

\begin{figure}[hbt]    
    \begin{picture}(200,200)(-50,10)
\put(0.,0.){
\includegraphics{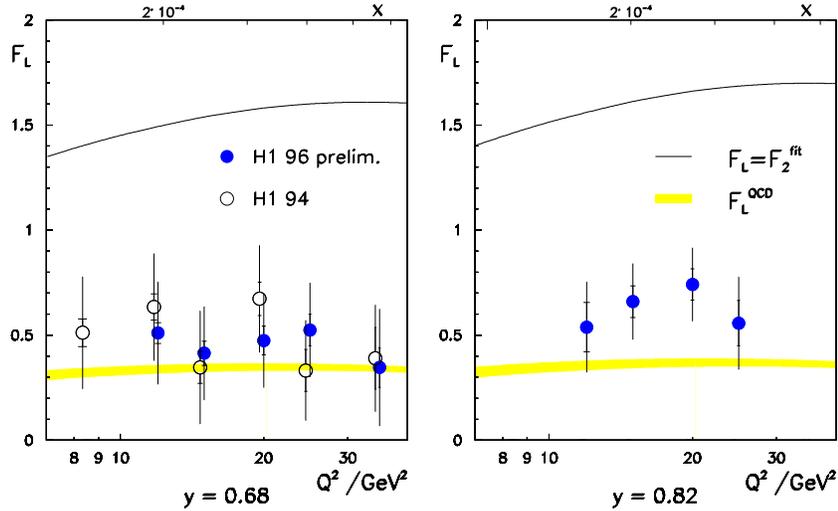}}
\end{picture}
   \caption{Longitudinal structure function 
$F_L =(F_2^{QCD fit} - \sigma / \kappa) \cdot Y_{+}/y^2$, where
$\kappa=(2 \pi \alpha^{2} \cdot Y_{+})/(Q^{4}x)$ and $Y_{+} = 1 + (1-y)^2$,
determined as function of $Q^2$
or $x=Q^2/sy$ for $y=0.68$ and $y=0.82$. 
The closed points represent
the preliminary H1 (1996) result while the open points are 
the published H1 (1994) data. The inner error bars are the
statistical error. The full error bars include the 
statistical and systematic errors added in quadrature. The error bands
represent the uncertainty of the calculation of $F_L$ using the gluon and 
quark  distributions as determined from a NLO QCD analysis of the
new H1 (1995, 1996) data for $y < 0.35$ and the fixed target experiment data.
The upper line defines the allowed upper limit of $F_L=F_2$ where $F_2$
is given by the QCD fit.}
\label{FL}
\end{figure}

In order to represent the cross section measurement 
 as a determination of $F_{L}$, a QCD fit was performed
 using the new H1 data (1995, 1996) only at low $y<0.35$ together with
 the BCDMS and NMC measurements.
The $F_L$ values shown in Figure \ref{FL} are calculated
from the $F_2$ values determined by this fit and the measured cross sections
in the high $y$ region. 
 The $F_2$ values from this fit agree within 2\% with the result 
 of the former fit~\cite{fl-h194} to the H1 (1994) and BCDMS data. 
For $y=0.68$ the new result on $F_{L}$ is in good agreement 
with the published 1994 data~\cite{fl-h194}  (open points). 
 
The total error of $F_L$ includes three different sources as discussed in
ref.~\cite{fl-h194}: the uncorrelated part of the systematic error
of the high $y$ cross section measurement, the systematic error of
the cross section correlated to the  error of the input data to
the QCD fit and  the error due to different assumptions inherent 
in the QCD fit. 
Out of the three error contributions the genuine high $y$ 
cross section uncertainty is the dominating one. 
The errors of $F_L$ at $y=0.82$ are 
 smaller than the errors at lower $y$ mainly due to the
 enhanced sensitivity to $F_L$ (factor $y^2$ in eq.~\ref{dsigma}).

The  calculation of $F_{L}$ in NLO QCD is 
given in Figure~\ref{FL} by a shaded band. 
The experimental uncertainty of this calculation is  about 6\%. 
The  data points are in agreement with QCD expectation,
however, they are systematically higher than expected (note that 
the points at given $y$ are highly correlated).
This tendency is also visible  in the cross section measurement at high 
 $y$ (Figure~\ref{sig}). 

A determination of $F_L$ at HERA which is free of any theoretical
assumptions is foreseen by measuring the $ep$ inclusive cross section
at different incident proton beam energies.

\subsection{The Very Low $Q^2$ Region}
With improved detectors in the backward region the ZEUS and H1 
data on $F_2$ cover now a range of $Q^2$ down to $\sim 0.1$~GeV${^2}$.
These data allow to study the transition from 
the region of perturbative QCD (DIS) to 
the photoproduction limit described by  Regge phenomenology.

\begin{figure}[hbt]    
    \begin{picture}(200,327)(-80,-15)
    \put(0,0){
\includegraphics{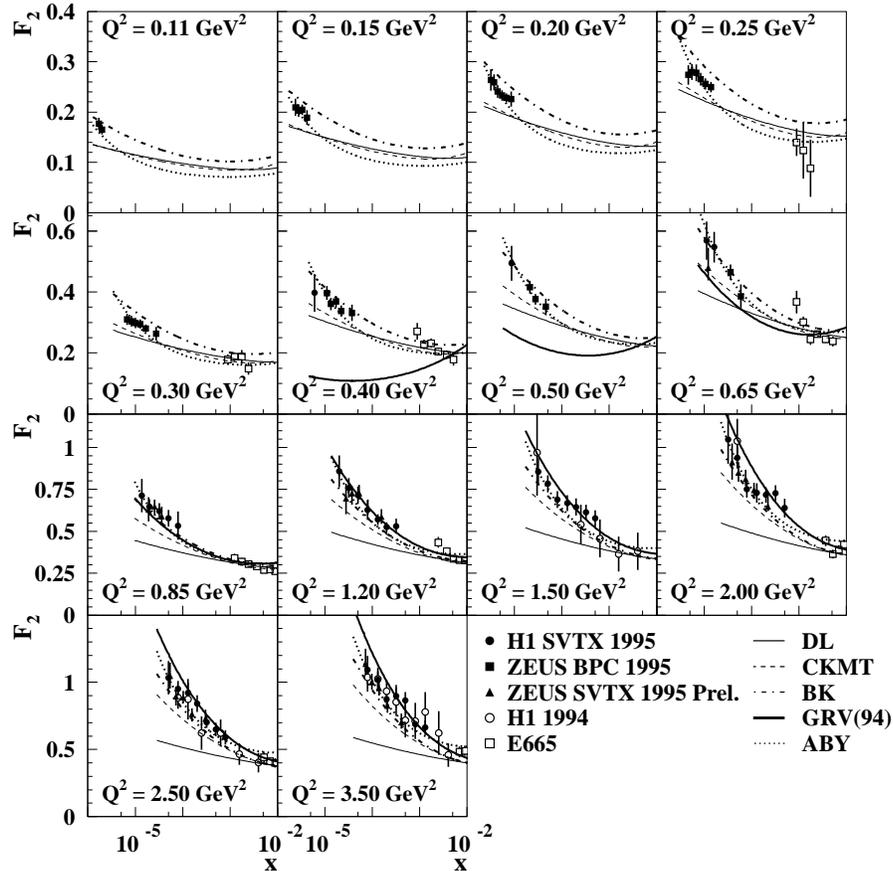}}
\end{picture}
\caption[]{Recent measurements of the proton structure function $F_2(x,Q^2)$
  in the low $Q^2$ region by H1 and ZEUS (full symbols), together with the 
  previous H1 measurements and results from the E665 experiment 
  (open symbols). Different models are compared with the data.}
\label{f2-lowq2}
\end{figure}

\begin{figure}[htb]
\begin{picture}(150,307)(0,0)
\put(40,50){\psfig{file=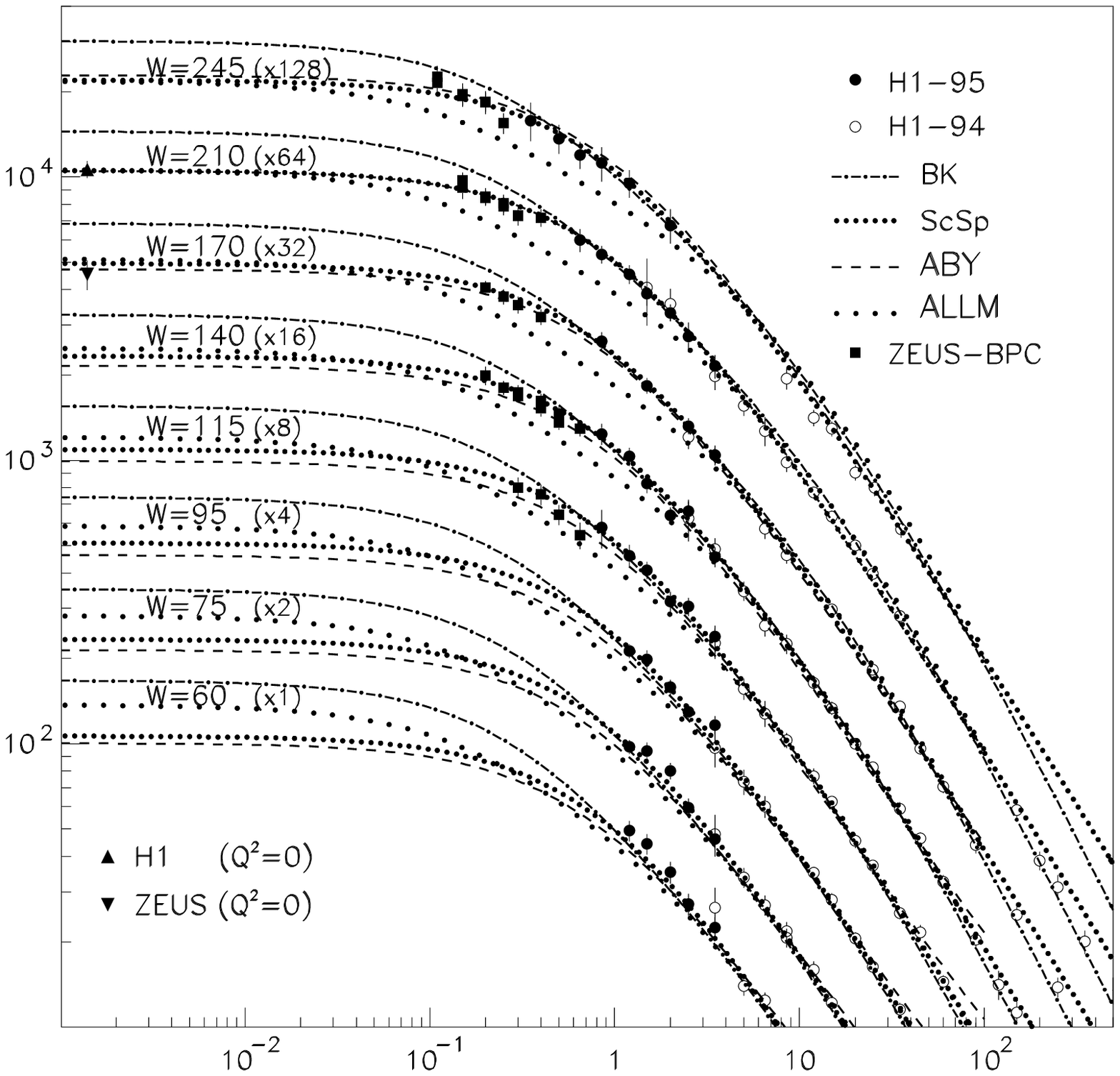,width=12.5cm}}
\put(25,280){{\large \boldmath $\sigma_{\gamma^* p}^{eff}$}} 
\put(25,260){{\large $\bf (\mu b)$}} 
\put(190,-3){{\large \boldmath  $Q^2$ \bf (GeV$^2$)}} 
\put( 87.,20.0){\large \boldmath $\wr$}
\put( 89, 20.0){\large \boldmath $\wr$}
\put( 87.,292){\large \boldmath $\wr$}
\put( 89, 292){\large \boldmath  $\wr$}
\end{picture}
\caption{  Measurement of the virtual photon-proton cross section
    $\sigma_{\gamma^*p}^{eff}$ by the HERA experiments 
    as a function of $Q^2$ at various values of $W$ (in GeV). 
    The photoproduction points as measured at HERA are also given.
    The cross sections for consecutive $W$ values are multiplied
    with the factors indicated in the figure (numbers in brackets).
    The curves represent different predictions for the transition 
    region from DIS to the photoproduction limit ($Q^2$=0).  
\label{sig-eff}}
\end{figure}

\begin{figure}[htbp]
\begin{picture}(150,145)(0.,0.)
\put(20.,-60.){\psfig{file=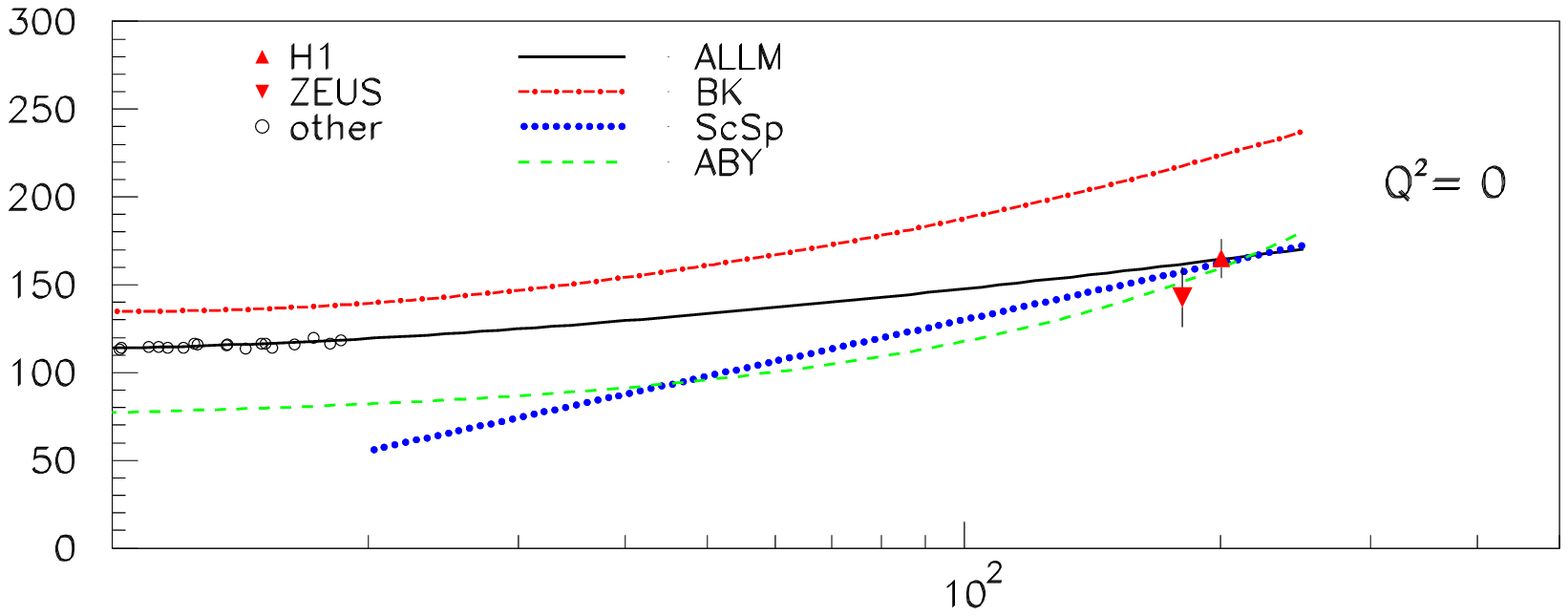,width=15.0cm}}
\put(10,120){{\large \boldmath $\sigma_{\gamma p}^{tot}$}} 
\put(10,100){{\large $\bf (\mu b)$}} 
\put(320,-8){{\large \boldmath  $W$ \bf (GeV)}} 
\end{picture}                                                               
\caption{Measurement of the real photon-proton total cross section
 $\sigma_{\gamma p}^{tot}$ as a function of $W$. 
\label{sig-gp}}
\end{figure}                                                                   

The low $Q^2$ $F_2$ results of the  HERA experiments are shown 
in Figure~\ref{f2-lowq2}. The ZEUS BPC data~\cite{f2-bpc} collected in 1995
cover $0.11 < Q^2 < 0.65$~GeV$^2$ and $2\cdot10^{-6} < x < 6\cdot10^{-5}$.
The total systematic error varies from 6\% to 11\% 
with an overall normalization uncertainty of 2.4\%.
Here, $R$ is assumed to be 0.
The H1 measurements~\cite{f2-h195} based on the shifted vertex running in 1995
cover $0.35 < Q^2 < 3.5$~GeV$^2$ and $x > 6\cdot10^{-6}$.
The total systematic and overall normalization errors are 5-10\% and 3\%,
respectively. 
The preliminary ZEUS shifted vertex data~\cite{f2-zeus95svx} from
the same period in 1995 cover the region $Q^2 > 0.65$~GeV$^2$.  
In the last two measurements $R$ is taken from ref.~\cite{fl-bk}.
In the region of overlap the results are in good agreement.
The data also show a smooth continuation from the fixed target measurements
towards the low $x$ region at HERA.
The rise of $F_2$ with decreasing  $x$ is very strong for
values of $Q^2 \ge 2$~GeV$^2$ but becomes less steep for smaller $Q^2$ values. 

Several parameterizations based on phenomenological models
are also shown in Figure~\ref{f2-lowq2}. Most of them use
ingredients both from Regge theory at low $Q^2$  
and from QCD when $Q^2$ is of the
order of 1~GeV$^2$ or larger (for a recent review, see e.g. \cite{levy}).

 Parameterizations motivated by Regge theory
 relate the structure function to
Reggeon exchange phenomena which successfully describe
the slow rise of the total cross section with energy
in hadron-hadron and $\gamma p$ interactions.
Using the ``bare'' instead of the ``effective'' pomeron intercept,
the CKMT~\cite{ckmt} parameterization
rises faster with $x$ compared to the DL~\cite{dola} calculations.
Regge inspired models  generally undershoot the data, except for the 
smallest $Q^2$ values where the calculations approach the data.

Also shown in Figure~\ref{f2-lowq2} are the predictions from 
the QCD-based GRV model~\cite{GRV}.
This model  assumes  that all parton distributions at
a very low  $Q^2_0= 0.34$ GeV$^2$
 have a valence like shape, i.e. vanish for $x\rightarrow 0$, 
and that the leading twist QCD evolution  equations can be used
to evolve the parton distributions from
this low $Q^2_0$ scale to larger $Q^2$ values.
Figure~\ref{f2-lowq2} shows that the GRV distributions describe the  data 
for $Q^2 \ge 1$ GeV$^2$, but systematically undershoot the data 
for $Q^2 < 1$~GeV$^2$.
  
In studies of the whole transition region starting from  $Q^2$=0
it is convenient to present the low~$Q^2$ data in terms of a
virtual photon-proton cross section~\cite{hand}.
The double differential $ep$ cross section, eq.~\ref{dsigma},
can be expressed via
the absorption cross sections for transverse and longitudinal virtual photons,
\begin{equation}                                                              
  \frac{d^2\sigma}{dx dQ^2}  
  = \Gamma [\sigma_T(x,Q^2) +  \epsilon(y)\sigma_L(x,Q^2)]
  \equiv \Gamma \sigma^{eff}_{\gamma^*p}(x,y,Q^2).
\label{sigeff}
\end{equation}
At small $x$ the following relations hold for 
 the flux factor $\Gamma$  and the photon polarization $\epsilon$: 
\newline
$\Gamma = \alpha (2-2y+y^2)/(2\pi Q^2x)$,
$\epsilon(y) = 2(1-y)/(2-2y+y^2)$.
The quantity $\sigma^{eff}_{\gamma^*p}$ is the effective measured 
virtual photon-proton
cross section for $ep$ collisions in the defined kinematic range, 
and can be determined from the data without assumptions on  $R$.
The total virtual photon-proton cross section, which is related to $F_2$ by
$\sigma^{tot}_{\gamma^*p}= \sigma_T + \sigma_L 
\simeq (4\pi^2\alpha/Q^2)F_2(x,Q^2)$,
depends only on $Q^2$ and $x$ (or~$W^2 = Q^2(1/x-1)$).
With the exception of the region of high $y$, where effects of $R$ are sizable,
$\sigma^{tot}_{\gamma^*p} \approx \sigma^{eff}_{\gamma^*p}$. 

Figure~\ref{sig-eff} shows the measured $\sigma^{eff}_{\gamma^*p}$ as a 
function of $Q^2$ for $W$ values above 60 GeV including the
photoproduction results~\cite{gp-h1,gp-zeus} from HERA.
The parameterization of Abramowicz et al. (ALLM~\cite{ALLM}) 
agrees well with the photoproduction data and the measurements 
at $Q^2 > 2$~GeV${^2}$,
but departs from the data around $Q^2 = 1$~GeV${^2}$. 

Also shown in Figure~\ref{sig-eff} are the predictions of the following models.
The model of Badelek and Kwiecinski (BK~\cite{BK})
combines the concepts of Generalized Vector Meson Dominance (GVMD) 
with dynamical parton models such as that of GRV.
A GVMD inspired approach has been proposed by
Schildknecht and Spiesberger (ScSp~\cite{ScSp}) to fit the low
and medium $Q^2$ HERA data up to $Q^2$ values of 350 GeV${^2}$.
A different approach to the low $Q^2$ behaviour in the transition 
region has been presented by Adel et al. (ABY~\cite{ABY}).
It assumes that perturbative  QCD  evolution
is applicable  to the lowest values of $Q^2$. 
All these models describe the HERA data rather well 
for $Q^2$ above  $\approx 0.1-0.4$~GeV${^2}$
but they fail
to describe the photoproduction $\sigma^{tot}_{\gamma p}$ data 
at low $W < 20$~GeV as it is shown in Figure~\ref{sig-gp}.
Thus, it turns out that there is no model found 
which is able to describe all existing data in a consistent way. 
However, there is a remarkable theoretical activity (e.g.
refs.~\cite{lowq2-levin,ALLM-new,lowq2-mrs})
in this field and progress can be expected soon. 

\section{Summary and Concluding Remarks}
With the final results by the NMC experiment on 
$\mu$$N\rightarrow\mu$$X$ deep inelastic scattering,
the present fixed target program for unpolarized DIS with charged
lepton beams is completed, providing very precise data sets 
which cover $0.2 < Q^2 < 260$~GeV$^2$ and $8\cdot10^{-4} < x < 0.9$.
The CCFR collaboration re-analyzed their $\nu$$Fe\rightarrow\mu$$X$
data and presented new results on $F_2$ and $xF_3$.
The data sets from the different fixed target experiments
and from HERA (though the region of overlap is small) are well consistent
apart from a difference of about 15\% 
in $F_2$ between NMC and CCFR at $x$=0.0125.
The updated $\alpha_S(M_Z^2)$ value 
0.119$\pm$0.002(exp.)$\pm$0.004(theory) 
from the re-analyzed CCFR data on 
$F_2$ and $xF_3$ is close to the LEP results.

The Fermilab experiments CDF, E866 and E706 presented new preliminary results
on the charge asymmetry in $W$ production, Drell-Yan $\mu$-pair production and 
prompt photon production, which provide more stringent constraints
on $u/d$ and $\bar{u}/\bar{d}$ ratios and the gluon density at high $x$.

The H1 and ZEUS experiments at HERA  
extended the previously accessible kinematic range
for $F_2$ up to very large values of $Q^2$ 
and down to very small values of $x$ covering the region
$0.1 < Q^2 < 5000$~GeV$^2$ and $2\cdot10^{-6} < x < 0.5$.
The conventional NLO DGLAP evolution describes  
the DIS data very well down to surprisingly  low
$Q^2$ values of about 1~GeV${^2}$.
Although there is no evidence for deviations from NLO DGLAP,
other approaches like the BFKL~\cite{BFKL} or CCFM~\cite{CCFM} evolutions 
(the latter combines both DGLAP and BFKL) can not be ruled out.
Another possible complication  at low $x$ is 
due to higher twist terms which may not be negligible even at 
$Q^2$ as large as 10~GeV$^2$ as discussed 
in refs.~\cite{levin,bartels}
and the success of DGLAP fits, perhaps, 
reflects the flexibility in choosing arbitrary functions
of $x$ for the input parton distributions.

The analysis of the HERA data in the framework of perturbative QCD
using NLO DGLAP evolution led to a determination of the gluon distribution
at low $x$. The gluon density
is rising steeply towards low $x$, as one could expect from 
the strong rise of $F_2$ with decreasing $x$.
H1 and ZEUS measured the charm contribution $F_2^{c\overline c}$ to 
the structure function of the proton which is
directly related to the gluon density at low $x$.
The charm contribution to $F_2$ is found to be $\approx$ 25\% and
the results are reasonably described 
in the three flavor number scheme 
with charm production via boson gluon fusion using
the gluon densities determined at HERA.
Another quantity related to the gluon density 
is the longitudinal proton structure function $F_L$.
H1 made an attempt to 
derive $F_L$ from the measured cross section at high $y$,
assuming that $F_2$ is given by a NLO QCD fit to the data at lower $y$.
The $F_L$ values are in broad agreement with QCD expectation,
however, systematically somewhat higher than expected.

The HERA measurements in the transition region from DIS to
the photoproduction limit have been confronted with different models.
There is no model found which is able to describe all
 existing data in a consistent way, but
further development of models is in progress.

\section{Acknowledgements}
It is a pleasure for me to thank all my colleagues from 
NMC, CCFR, CDF, E866, E706, ZEUS and H1 for providing information.
Special thanks to the H1 people working on structure function
measurements for their tremendous efforts to meet the date with new results.
I wish to thank R.~Eichler for encouragement and A. Wagner for support,
M. Ryskin for long and fruitful discussions,
J. Bartels, Yu. Dokshitzer, R. Engel, E. Levin, A.D. Martin, A. Prinias,
A. Quadt, J. Smith, B.~Surrow
for helpful conversations and
J. Gayler and A. Zhokin for help in the preparation of the manuscript.
My thanks to A. Caldwell, A. De Roeck, M. Klein and A. Levy
for careful reading of the final version of the manuscript.

\section*{References}

\end{document}